\documentclass[conference]{IEEEtran}
\IEEEoverridecommandlockouts
\usepackage{cite}
\usepackage{amsmath,amssymb,amsfonts}
\usepackage{algorithmic}
\usepackage{graphicx}
\usepackage{textcomp}
\usepackage{float}
\usepackage{etoolbox}
\usepackage{xcolor}
\usepackage[caption=false]{subfig}
\usepackage{filecontents}
\usepackage{ragged2e}

\makeatletter
\patchcmd{\@makecaption}
  {\scshape}
  {}
  {}
\makeatletter
\patchcmd{\@makecaption}
  {\\}
  {.\ }
  {}
  {}
\makeatother
\def\tablename{Table}
\def\BibTeX{{\rm B\kern-.05em{\sc i\kern-.025em b}\kern-.08em
    T\kern-.1667em\lower.7ex\hbox{E}\kern-.125emX}}
\begin{document}

\title{Modality Fusion Network and Personalized Attention in Momentary Stress Detection in the Wild
\thanks{This work is supported by NSF \#2047296 and \#1840167}
}

\author{\IEEEauthorblockN{Han Yu}
\IEEEauthorblockA{\textit{Rice University} \\
Houston, USA \\
han.yu@rice.edu}
\and
\IEEEauthorblockN{Thomas Vaessen}
\IEEEauthorblockA{\textit{KU Leuven	} \\
Leuven, Belgium \\
thomas.vaessen@kuleuven.be}
\and
\IEEEauthorblockN{Inez Myin-Germeys}
\IEEEauthorblockA{\textit{KU Leuven} \\
Leuven, Belgium \\
inez.germeys@kuleuven.be}
\and
\IEEEauthorblockN{Akane Sano}
\IEEEauthorblockA{\textit{Rice University}\\
Houston, USA \\
akane.sano@rice.edu}
}

\maketitle

\begin{abstract}
Multimodal wearable physiological data in daily life settings have been used to estimate self-reported stress labels.
However, missing data modalities in data collection make it challenging to leverage all the collected samples. Besides, heterogeneous sensor data and labels among individuals add challenges in building robust stress detection models. In this paper, we proposed a modality fusion network (MFN) to train models and infer self-reported binary stress labels under both complete and incomplete modality condition. In addition, we applied a personalized attention (PA) strategy to leverage personalized representation along with the generalized one-size-fits-all model. We evaluated our methods on a multimodal wearable sensor dataset (N=41) including galvanic skin response (GSR) and electrocardiogram (ECG). Compared to the baseline method using the samples with complete modalities, the performance of the MFN improved by 1.6\% in f1-scores. On the other hand, the proposed PA strategy showed  a 2.3\% higher stress detection f1-score and approximately up to 70\% reduction in personalized model parameter size (9.1 MB) compared to the previous state-of-the-art transfer learning strategy (29.3 MB).
\end{abstract}


\section{Introduction}
Psycho-social stress is a common phenomenon that can benefit people under certain circumstances and increase resilience to future stressors. Especially exposure to moderate stress levels can be beneficial as it can prepare an organism to deal with future challenges \cite{dhabhar2014effects}. On the other hand, stress has also been associated with an increased risk for a large number of somatic and mental illnesses \cite{aschbacher2013good}. Prolonged exposure to a stressful environment increases risks for cardiovascular health issues \cite{kario2003disasters} and suppresses the human immune system \cite{khansari1990effects}. Effectively detecting moments of stress in real life may provide opportunities to help individuals manage their stress early on to promote resilience and wellbeing.

In recent years, the rapid development of mobile sensors and machine learning algorithms have brought the opportunity to measure momentary human stress using passive multimodal physiological sensor data such as heart rate, sleep activity, galvanic skin response (GSR), and blood volume pulse (BVP) \cite{sanchez2017towards, gjoreski2017monitoring, shi-stress}. These prior works have shown promising results and have revealed the feasibility of using multimodal physiological data to monitor human stress. However, we hypothesize addressing the following two issues can provide further improvement. 

\textbf{Missing modalities} are common in real-world multimodal sensor data since it is challenging to ensure that all sensors function continuously throughout data collection period. The studies mentioned above only used data with all complete modalities of features to maintain the quality of data used for stress modeling; however, as one of the previous methods, discarding samples with incomplete modalities leads to information loss. For example, in a case that data from one sensor (e.g., a chest-worn sensor) were not collected as expected due to various reasons such as sensor malfunction or battery drainage, nevertheless, data from wrist-worn sensors and participants' subjective stress labels were collected, the stress label can be still estimated only using the data from the wrist-worn sensor rather than discarding those samples. Deep learning methods with missing modalities has been developed in computer vision and natural language processing \cite{ma2021smil, pandey2017variational, wang2018lrmm}.
Some prior studies also developed deep learning methods to impute the missing modalities in human emotion studies \cite{jaques2017multimodal, zhang2019ssim, peralta2021data, bucur2018early, mittal2020m3er}. However, the data reconstruction errors from models such as principle component analysis and auto-encoder in these methods would introduce biases into the imputed modalities, especially when the entire modalities were missing in the input samples. In this work, instead of pre-imputing the physiological data, we proposed an end-to-end modality fusion network (MFN), which models the data from two sensors under the either complete or incomplete modalities situation.

\textbf{Heterogeneous sensor data and labels among individuals} are another common problem that affects the robustness of stress detection models. Heterogeneity exists in both collected sensors data and self-reported stress labels. For instance, physiological measurements show the differences in heart rates and body acceleration for different persons; meanwhile, the perception of stress varies from person to person. Thus, it is difficult to build a one-size-fits-all model that estimates stress status accurately for every participant. Some prior studies have leveraged individual heterogeneity in modeling stress \cite{can2019continuous, Previous-cla, saeed2018model, yu2020passive}. Taylor \textit{et al.} clustered participants' data based on the genders and personality information and applied a multi-task learning strategy to predict participants' wellbeing labels include stress as different tasks \cite{Previous-cla}. Yu \textit{et al.} transferred a one-size-fits-all model on each subject's data and observed an improved overall stress prediction accuracy \cite{yu2020passive}. Nevertheless, these methods cannot fully solve issues in heavy computational complexity and weak model adaptability, which prevent these models from being used in real-world applications and benefiting people. For example, if 25 different groups of participants (25 output branches) are found and used in the above multi-task learning model, only one branch is available for a sample during training and prediction. This mechanism causes the remaining 24 branches to perform unnecessary computations. In this work, we designed a personalized attention mechanism, which provided over 77\% f1-score in personalized stress estimation with low computational cost and flexible adaptability. 

To summarize our contributions:
\begin{itemize}
    \item We proposed an MFN model structure to estimate binary momentary stress labels even with missing data modalities of wearable sensor data.
    \item We designed a personalized attention strategy to model momentary stress with individual heterogeneity. This method provided more accurate stress estimation results with less computation cost and smaller model parameters size compared to the previous methods.
\end{itemize}

\section{Related Work}

With the development of mobile phones and wearable devices, accessing users' physiological and behavioral data in daily life settings has become popular. Machine learning enables us to develop models to learn patterns from data samples and has already benefited ubiquitous computing applications. Multimodal data from wearable sensors, mobile phones, and other smart devices have been widely used with machine learning in estimating momentary stress levels \cite{sanchez2017towards, gjoreski2017monitoring, shi-stress}.
Shi \textit{et al.} collected 22 subjects' electrocardiogram (ECG), GSR, respiration (RIP), and skin temperature (ST) data using wearable sensors \cite{shi-stress}. Each subject in the study was exposed to a protocol, including four stressors and six rest periods, and stress labels were collected before and after each stressor/rest period through interviews. Then the authors proposed a personalized SVM algorithm to classify binary stress labels (low/high), which provided 0.68 precision with 0.80 recall. Some studies have also compared the effects of using different sensor modalities on stress estimation models \cite{can2019continuous, umematsu2019improving, Previous-reg2}. For example, Can \textit{et al.} compared the performance of stress estimation using different sensor devices and various signal types \cite{can2019continuous}. They used two types of sensors: the Empatica E4 sensor and the Samsung Gear S-S2, and their results showed that the combined modalities of heart rate with electrodermal activity fitted the model with the highest accuracy. 
These works inspired us with the important insight that we can estimate stress by using different modalities separately while using different modalities in combination often yields better estimation results. 

However, the modality missing issue is inevitable in multimodal data collection. To ease the loss from incomplete modalities, researchers have proposed various deep learning network structures \cite{ma2021smil, pandey2017variational, wang2018lrmm, cai2018deep}. For example, Ma \textit{et al.} proposed a multimodal learning with severely missing modality model, which used Bayesian meta-learning to reconstruct severely missing modalities with valid modalities \cite{ma2021smil}. They evaluated the proposed method on the MM-IMDb dataset (image + text) and CMU-MOSI dataset (image + audio), and showed that the model achieved higher performances on both datasets than the baseline models.
Researchers also designed methods of imputing missing values in multimodal physiological dataset \cite{jaques2017multimodal, zhang2019ssim, peralta2021data}. Jaques \textit{et al.} proposed an auto-encoder structure (MMAE) that imputed missing values in multimodal human physiological and behavioural data \cite{jaques2017multimodal}. The MMAE methods outperformed the baseline principal component analysis method in data reconstruction root mean squared error (RMSE) metrics. Zhang \textit{et al.} proposed a sequence-to-sequence imputation model (SSIM) framework to impute missing values for multimodal time-series sensor data \cite{zhang2019ssim}. Their evaluated model achieved up to 69.2\% improvement in the reconstruction RMSE score compared to a baseline matrix factorization method. To the best of our knowledge, there is no previously designed end-to-end model that adaptively used different physiological modalities and their combinations to perceive stress labels without reconstructing the missing modalities.

Since the subjectively perceived stress of participants is related to individually heterogeneous sensor data and labels, the model personalization has also been a topic of interest for researchers \cite{can2019continuous, Previous-cla, saeed2018model, yu2020passive}. For example, Taylor \textit{et al.} \cite{Previous-cla} applied a K-Means algorithm to cluster participants into different groups by their personality survey and gender information. Using the group information, they built diverse output branches in the neural network for different groups of participants to implement multi-task learning. 
These works all concluded that personalized models achieved better performance in human stress estimation than generalized models. However, there were drawbacks in these previous works. For example, in multi-task learning models, high computational complexity and low adaptability for new participants to the pretrained models have been a existing problem. If there is discrepancy in data characteristics between new participants and the participants used for training models, the multi-task learning model needs to be re-trained. On the other hand, although the transfer learning strategy adapts the model to new samples, storing the parameters of the transfer learning model might not be efficient in real-world applications. For each individual, parameters of the fine-tuned layers as well as the whole model might need to be stored on each edge device, which is detrimental to the system's space complexity.

\section{Data Set}
\label{dataset}
In this study, wearable sensor and self-report data were collected from 41 healthy participants (36 females and 5 males). The average age of participants was 24.5 years old, with a standard deviation of 3.0 years.

\subsection{Wearable sensor data}

Two types of wearable sensors were used for data collection \cite{smets2018towards}. One was a wrist-worn device (Chillband, IMEC, Belgium) designed for the measurement of galvanic skin response (GSR), which was sampled at 256 Hz. Participants wore the sensor for the entire testing period, but could take it off during the night and while taking a shower or during vigorous activities. The second sensor was a chest patch (Health Patch, IMEC, Belgium) to measure ECG. It contains a sensor node designed to monitor ECG at 256 Hz throughout the study period. Participants could remove the patch while showering or before doing intense exercises.

Data from both chest and wrist wearable sensors were sorted based on their timestamps, and a set of 16 features was computed. Table \ref{features_table} shows the features computed from ECG and GSR signals with 5-minute sliding windows with 4-minute overlapping segments \cite{smets2018towards}. 

\begin{table}
\centering
\caption{Features. ECG: electrocardiogram, GSR: galvanic skin response}
\vspace{-0.2cm}
\begin{tabular}{c|p{5cm}|c}
\label{features_table}No. & Feature                                                                                                                           & Source                                                 \\ \hline
1                         & Mean heart rate                                                                                                              & ECG                                                    \\
2                         & Standard deviation of heart rate variability's R-R intervals                                                                      & ECG                                                    \\
3                         & Root mean square of successive R-R differences                                                                                    & ECG                                                    \\
4                         & Low frequency signal (power in the 0.04-0.15 Hz band)                                                                             & ECG                                                    \\
5                         & High frequency signal (power in the 0.15-0.40 Hz band)                                                                            & ECG                                                    \\
6                         & Ratio of low and high frequency                                                                                                   & ECG                                                    \\
7                         & Ratio of very low (0.0033 - 0.04 Hz) and low frequency                                                                            & ECG                                                    \\
8                         & Heart rate cycle                                                                                                                  & ECG                                                    \\ \hline
9                         & GSR level - average GSR                                                                                                             & GSR                                                     \\
10                        & Phasic GSR - signal power of the phasic GSR signal (0.16-2.1 Hz)                                                                    & GSR                                                     \\
11                        & GSR response rate - number of GSR responses in window divided by the totally length of the window (i.e. responses per second)       & GSR                                                     \\
12                        & GSR second difference - signal power in second difference from the GSR signal                                                       & GSR                                                     \\
13                        & GSR response - number of GSR responses                                                                                              & GSR                                                     \\
14                        & GSR magnitude - the sum of the magnitudes of GSR responses                                                                          & GSR                                                     \\
15                        & GSR duration - the sum of the magnitudes of GSR responses                                                                           & GSR                                                     \\
16                        & GSR area - the sum of the area of GSR responses in seconds. The area is defined using the triangular method (1/2 x GSR magnitude x GSR duration) & GSR                                                     \\ \hline
\end{tabular}
\vspace{-0.6cm}
\end{table}

\vspace{-0.1cm}
\subsection{Momentary Stress Labels}
In addition to the physiological data collected by sensors, participants received notifications on their mobile phones to report their momentary stress levels 10 times at random timing per day for eight consecutive days. In total, 2494 stress labels were collected across all participants (80\% compliance). The stress scale ranged from 0 ("not at all") to 6 ("very"). The portions of each stress level's labels were 44.8\%, 17.8\%, 13.4\%, 11.2\%, 3.4\%, and 1.0\% from no stress at all to the highest stress level, respectively.  


\section{Methods}
This section introduces the details of our methods in building a stress detection system, including data pre-processing, baseline self-attention network (SAN), MFN, and PA.

\vspace{-0.1cm}
\subsection{Data Pre-processing}
\label{data_preprocessing}
We split each participant's data into 60-minute time windows, and any time windows with missing data points were omitted. There are 1123 valid sequences from GSR features, whereas 2107 sequences with ECG features, respectively. After merging these two modalities, only 938 sequential samples can be used to build the stress detection model using both modalities.

Stress labels were divided by the participants' self-reported stress status. We coded the stress status 0 as non-stressed labels and the rest 1-6 stess status stressed label to build a binary classifier. The ratio of the number of non-stressed labels and stressed labels was 45\% to 55\%.

\vspace{-0.1cm}
\subsection{Self-Attention Network}
\label{sec:san}
In recent years, time series learning has attracted attention from researchers. For example, as an essential member in the neural network family, recurrent neural network (RNN) was used in stock price prediction\cite{selvin2017stock}, machine translation\cite{cho2014learning}, and music generation\cite{goel2014polyphonic}. However, RNN has drawbacks such as high computational complexity and weak long-term dependency learning ability. To address long-term dependency issues, an attention mechanism with RNN has been used and achieved improvements in different areas \cite{wang2016survey, hori2017advances}. Further, the Transformer \cite{vaswani2017attention}, a structure that uses only the attention mechanism without the computational-heavy RNN unit has been developed. The Transformer achieved state-of-the-art performances in multiple applications \cite{vaswani2017attention, so2019evolved}. In this study, we applied the self-attention mechanism in the Transformer to extract representations from the 60-minute input sequential physiological data (60 steps x 1 min). The model outputted the results of stress detection in non-stressed/stressed binary labels.

\begin{figure}
	\centering
	\includegraphics[scale=0.2]{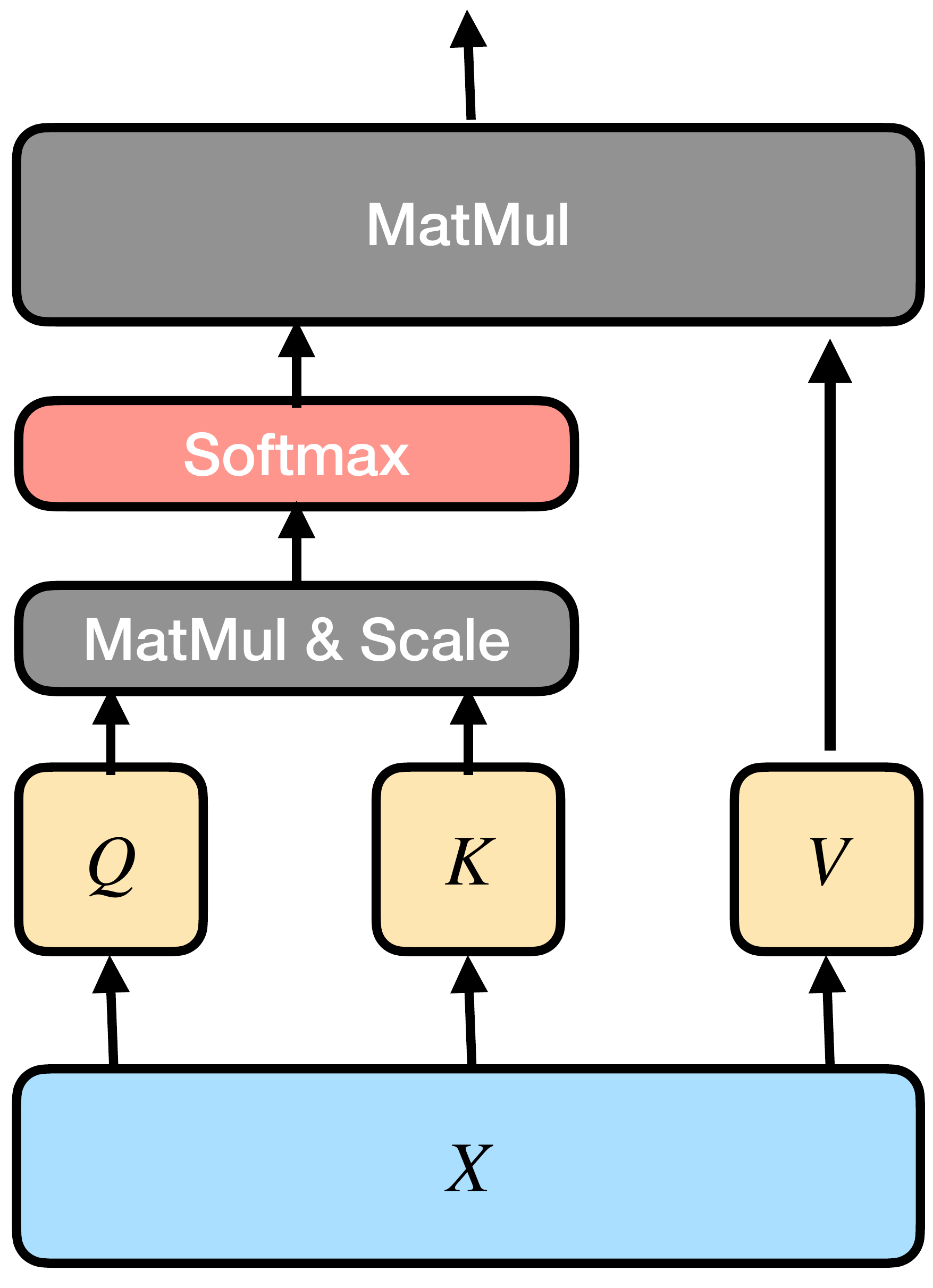}
	\vspace{-0.3cm}
	\caption{The self-attention mechanism. $X$ is the data input, $Q$, $K$, and $V$ are the projection matrices of $X$.}
	\label{attention_structure}
	\vspace{-0.5cm}
\end{figure}

According to the \cite{vaswani2017attention}, self-attention (Figure \ref{attention_structure}), which is an attention mechanism relating different positions of a single sequence to compute the representations, 
can be defined as:
\begin{equation}
\vspace{-0.1cm}
    \text{Attention}(Q, K, V) = softmax(\frac{QK^{\intercal}}{\sqrt{d_k}})V
    \label{eq:self-attention}
\end{equation}
where $Q$, $K$, and $V$ are the linear mappings from the original sequence $X$, and $d_k$ represents the dimension of matrix $Q$ and $K$. The above equation can be understood as the embedding of matrix $Q$ with the references of $V$. The scale factor $\sqrt{d_k}$ regulates the parameters and avoids the vanishing issues of the \textit{softmax} gradient when the inner product of $Q$ and $K$ is large.

\begin{figure}
	\centering
	\includegraphics[scale=0.2]{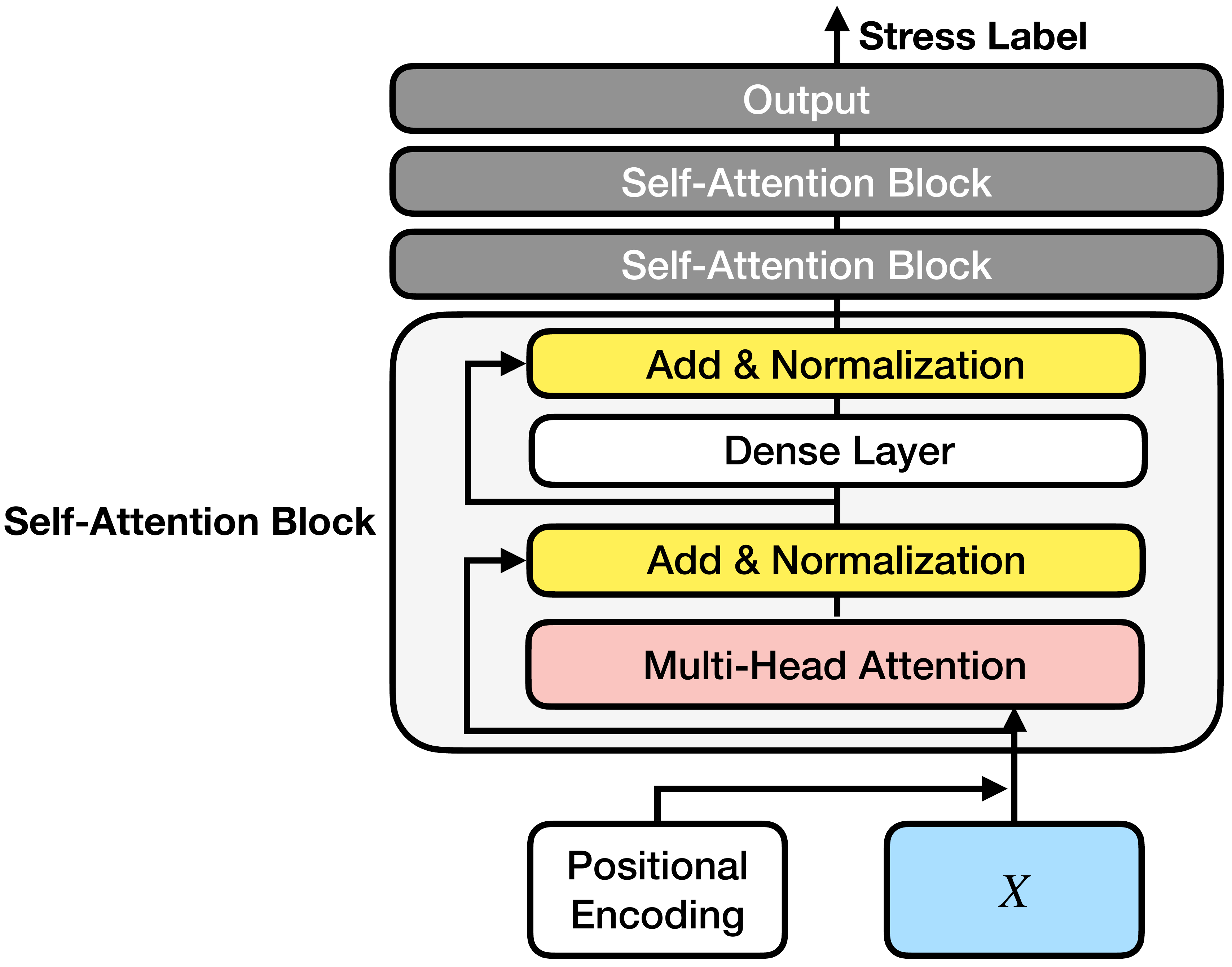}
	\vspace{-0.3cm}
	\caption{The structure of the self-attention network (SAN) in this study. $X$ is the input sequential physiological data. Four layers including one multi-head attention layer, one dense layer with two add \& normalization layers form a self-attention block. After positional encoding, the model passes input data through 3 self-attention blocks and outputs the stress estimation result.}
	\label{attention_network_structure}
	\vspace{-0.5cm}
\end{figure}

As stated in the original work \cite{vaswani2017attention}, the single self-attention mechanism may not be robust enough. Therefore, we decided to use a multi-headed attention mechanism, which is the concatenation of multiple self-attention layers to improve this problem. The multi-head attention can be referred to:
\begin{equation}
    \text{MultiHead}(Q, K, V) = \text{Concat}(head_1, head_2, ... , head_h)
\end{equation}
\begin{equation}
    head_i = \text{Attention}(Q_i, K_i, V_i)
\end{equation}
where $h$ is the number of the concatenated self-attention layers in a multi-head attention layer. Intuitively, this algorithm calculates the self-attention using equation \ref{eq:self-attention} for $h$ times and concatenates the results from all calculations. In this work, we tuned the $h$ as 4. As shown in Figure \ref{attention_network_structure}, we defined a self-attention block as an in-series combination of a multi-head attention layer, two "Add \& Normalization" layers, and one dense layer. The "Add \& Normalization" layers residually connected \cite{he2016deep} the input and multi-head attention to avoid the over-fitting issue, then parameters were also normalized in this layer to speed up the model optimization. 
A dense layer was also included in a self-attention block to feed forward information extracted by the multi-head attention. As shown in Figure \ref{attention_network_structure}, to make the model deeper and gain higher generalizability, we applied three such self-attention blocks in series to form the baseline SAN model. Besides, since the self-attention mechanism cannot extract the temporal representation, we applied a sinusoidal positional encoding method as in \cite{vaswani2017attention}, which generated the sequences of tokens to help the model recognize the relative positions of each time step in the input sequences.

\subsection{Modality Fusion Network}
\label{sec:MFN}
\begin{figure}
	\centering
	\includegraphics[scale=0.2]{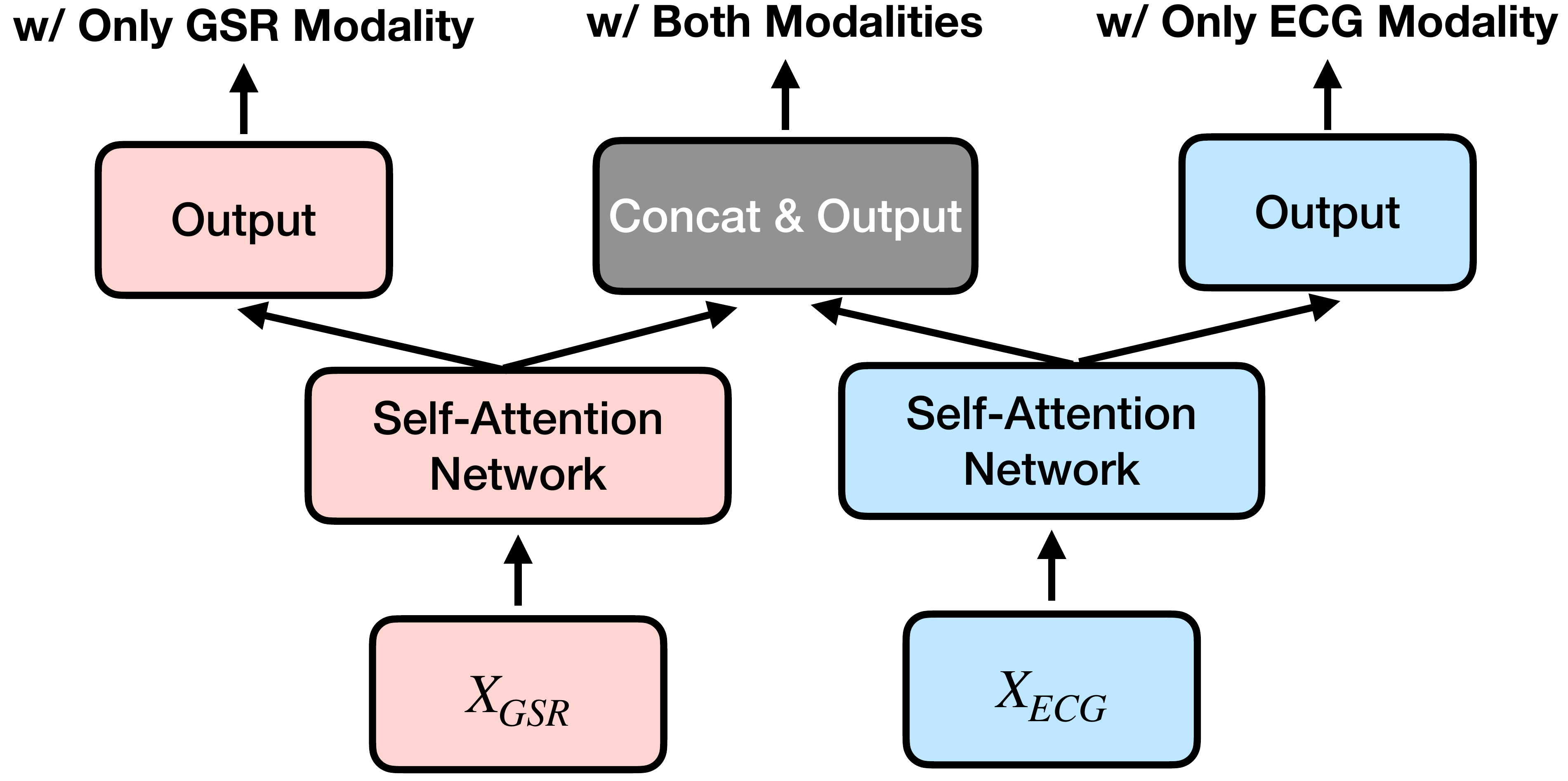}
	\vspace{-0.2cm}
	\caption{The structure of the modality fusion network. $X_{GSR}$ and $X_{ECG}$ are the input GSR and ECG data sequences, respectively.}
	\label{mfn}
	\vspace{-0.5cm}
\end{figure}

As the data statistics was shown in section \ref{data_preprocessing}, the number of valid ECG sequential samples was about 2 -- 2.5 times larger than the number of the valid GSR samples and the number of samples with complete 2 modalities. Thus, to fully leverage the collected data, it is crucial to develop a model that can handle both incomplete and complete modalities. In this work, we proposed an MFN model shown in Figure \ref{mfn}. This MFN model adaptively leverages  multimodal data to model stress patterns as explained in the next paragraph.

\textbf{Training:} Training procedures of the MFN network adapt to the missing condition of modalities. For example, if only the ECG data is valid in samples, we inputted randomly generated values into the GSR branch. However, the loss weights of both the GSR branch and the concatenated branch were set as 0. In this case, the model was only optimized through the ECG branch (blue in Figure \ref{mfn}). Similarly, the exact optimization mechanism was also applied in the case where only GSR data was available. The model would be optimized through all three output paths on the samples with both modalities. The loss function of MFN is written as:
\vspace{-0.1cm}
\begin{equation}
    \boldsymbol{loss}_{MFN} = I_G\cdot\boldsymbol{loss}_{GSR} + I_{E}\cdot\boldsymbol{loss}_{ECG} + I_{GE}\cdot\boldsymbol{loss}_{concat}
\end{equation}
Where $I_G$, $I_E$, and $I_{GE}$ are the indicator functions for missing modalities. For instance, $I_G$ and $I_{GE}$ are both 0 when the GSR data is missing.

\textbf{Stress Inference:} For inferring stress labels from MFN, we selected the outputs according to input data modality availability. If there was no missing modality, the concatenated output cell results were used as the final estimation. Otherwise, we selected the results from the branch with available modalities as the output of the model.

\subsection{Personalized Attention}
Learning personalized models from heterogeneous data with individual differences is still a challenging topic in human behavioral modeling. For example, in a recommendation system, personalized differences need to be considered to provide proper recommendations to users. 

\begin{figure}
	\centering
	\includegraphics[scale=0.2]{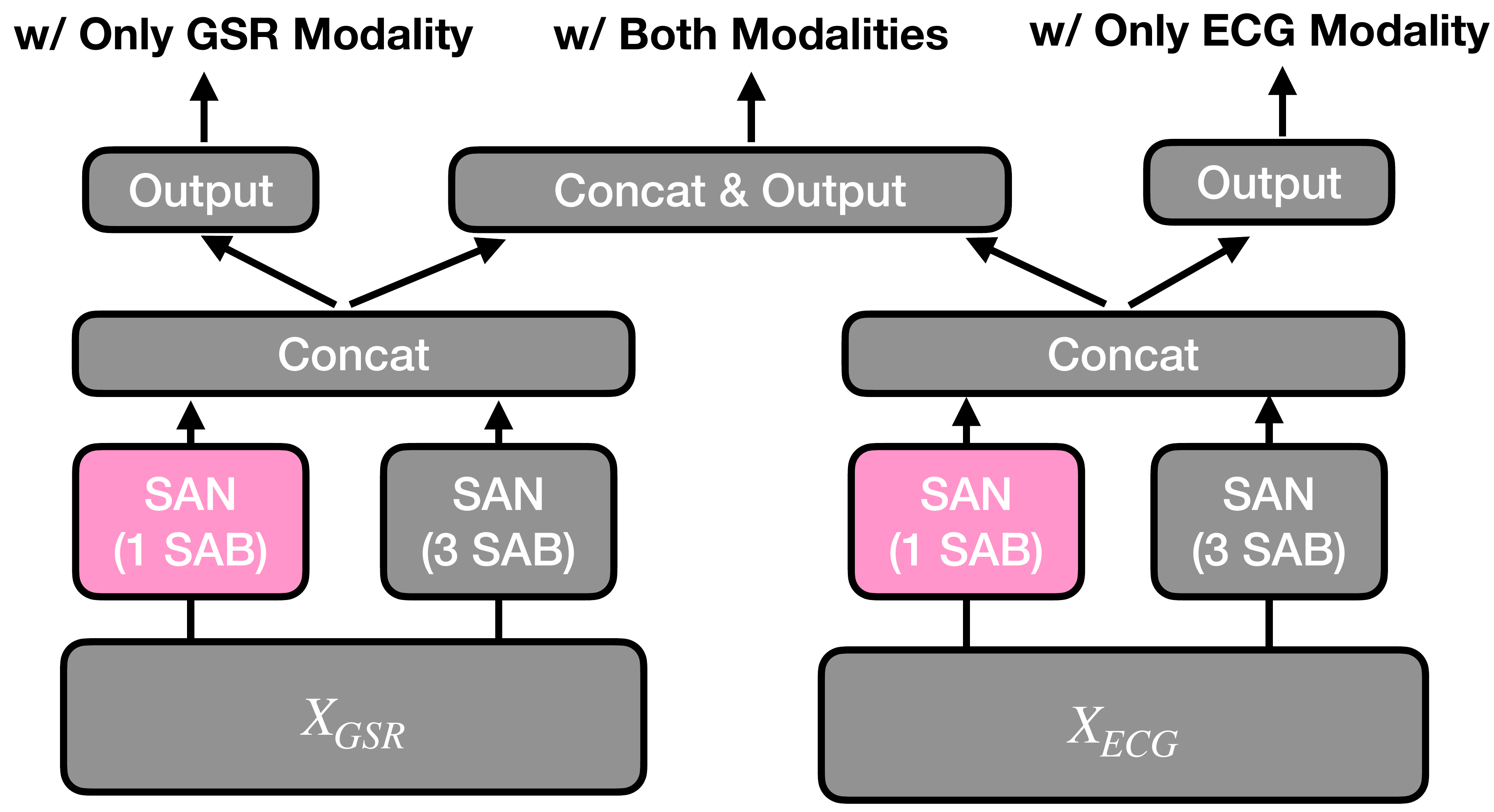}
	\vspace{-0.2cm}
	\caption{The structure of the modality fusion network-integrated personalized attention framework. $X_{GSR}$ and $X_{ECG}$ are the input GSR and ECG data sequences. SAN represents the self-attention network. Three self-attention blocks (SAB) in Section \ref{sec:san} are used in the generalized branch as SAN (3 SAB); while only one SAB is applied in the pink-highlighted personalized branch as SAN (1 SAB).}
	\label{personalized-attention}
	\vspace{-0.5cm}
\end{figure}

In this study, inspired by the method in \cite{liu2019nrpa}, we designed a personalized attention framework to combine the generalized information from a one-size-fits-all MFN model with the individual representations learned from tiny personalized sub-branches. Figure \ref{personalized-attention} shows the structure of the MFN-integrated PA framework designed in this work. In both the GSR and ECG branches in MFN, we added sub-branches using SAN with only one self-attention block as personalized attention layers. As a pre-preparation step for this structure, a one-size-fits-all MFN (section \ref{sec:MFN}) model was fitted with all participants' data in training set. Then, we fixed the parameters in the generalized MFN and constructed a personalized attention branch for individuals to learn personalized representations. The personalized branch structure can be considered a small MFN, with only one self-attention block instead of three blocks in a standard MFN. During the model training process, only parameters in the personalized attention layer and the concatenate and output layer are trainable. To fit the data for each participant, we initialized the parameters in the personalized attention layer for different individuals so that the network learned parameters for each participant that were specific to that participant only. 

\subsection{Focal Loss}
In this work, as shown in Section \ref{dataset}, the stress labels are not uniformly distributed. To avoid the issues from data imbalance, i.e., the classifier was dominated by the major class, caused by imbalanced data sets, we applied a focal loss \cite{lin2017focal} as the loss function of our model. The focal loss is written as:
\begin{equation}
    FL(\mathnormal{p}_t) = - (1 - \mathnormal{p}_t)^{\gamma}\log{\mathnormal{p}_t}
\end{equation}
where $\mathnormal{p}_t$ is a raw sigmoid output from the model, and $\gamma$ is the focusing parameter. The purpose of using this loss function is to make the model more focused on hard-to-classify samples during training by reducing the weights of easy-to-classify samples. The $\gamma$ value we used in this study was 2.

\section{Experiments}
Considering that human behavior patterns or stress perceptions might change over time, we referred to the study timeline for splitting the participants' data to simulate the realistic conditions. We sorted the reported stress labels for each participant by the time of reporting, and then we selected the first 70\% as our training set and the latter 30\% as our test set. With this setup, we can ensure that there is not any time series overlap between the training and test sets and simulate using a model fitted by past data to infer future unseen data. Furthermore, to avoid biases from different model initialization that leads to the lack of rigor in comparing evaluation results, we repeated the training and testing process five times.

After splitting the dataset, the numbers of positive and negative samples were not equal in the training and the test set. Therefore, considering the bias of accuracy rate caused by the imbalance problem, we chose the f1-score as the metric in our model performance evaluations.

The following experiments were conducted to evaluate our proposed methods.

\subsection{Modality Fusion Evaluation}
As described in sections \ref{dataset}, we have two different data modalities, ECG and GSR. There are three different possible scenarios of data availability: i) ECG data only, ii) GSR data only, and iii) both. In this experiment, we used SAN and MFN in 3 scenarios (i), (ii), and (iii) and compared their stress detection performances. As SAN requires fixed input dimensions, we used 3 SAN models to model (i), (ii), and (iii), respectively; while one MFN model was used to estimate stress labels for all 3 scenarios. Further, a paired t-test was applied to compare the performances of SAN and MFN.

\subsection{Model Personalization}
For evaluating the personalized model, we used the generalized MFN model as a baseline. Then we compared the personalized MFN model with the baseline to assess improvement from personalization. We also applied the previous transfer learning strategy \cite{yu2020passive} on this dataset for comparing the personalized MFN and the previous state-of-the-art transfer learning method. To implement the transfer learning method on our dataset, we first trained a one-size-fits-all MFN model based on all subjects' training data. Using the parameters of the generalized model as the initialization, we fixed the parameters in the first two blocks of the MFN and fine-tuned the last MFN block parameters for each participant. In this experiment, in addition to calculate f1-score of the stress detection results, we also evaluated the model size for each participant as another criterion. Such a criterion is necessary because smaller models could be easier to adapt to the edge devices in real-world applications. We used an ANOVA one-way test to compare the one-size-fits-all baseline model and two personalized models. Also, we applied a paired t-test to compare f1-score performances between the transfer learning strategy and PA framework.
\section{Results}
\subsection{Modality Fusion Evaluation}
Table \ref{res:mfn} shows the evaluation results of using SAN and MFN with different modalities of data. Our proposed MFN showed higher f1-scores than each of the SAN models. 
The p-values of the statistic tests were all smaller than 0.01, which indicated that our proposed MFN model performed statistically significantly better than the baseline SAN models in the f1-scores. 

\begin{table}
\caption{Stress detection model performances (f1-score) using various modalities. SAN represents the self-attention network. Three SAN models were applied to samples with only GSR or ECG modality, or both modalities. MFN represents the modality fusion network we proposed. Bold indicates statistically significant differences between MFN and SAN (paired t-test, p-value $<$ 0.01).}
\centering
\label{res:mfn}
\vspace{-0.2cm}
\begin{tabular}{c|c|c|c|}
\cline{2-4}
                                                            & GSR Only                   & ECG Only                   & GSR + ECG                  \\ \hline
\multicolumn{1}{|c|}{SAN} & 0.629 $\pm$ 0.007          & 0.651 $\pm$ 0.003          & 0.677 $\pm$ 0.004          \\ \hline
\multicolumn{1}{|c|}{MFN}                                   & \textbf{0.646 $\pm$ 0.005} & \textbf{0.660 $\pm$ 0.005} & \textbf{0.693 $\pm$ 0.006} \\ \hline
\end{tabular}
\vspace{-0.4cm}
\end{table}
\subsection{Model Personalization}
Table \ref{res:pa} shows the results of the generalized MFN model and the personalized models, including transfer learning and personalized attention strategies, respectively. We observed improvements in the f1-score of the personalized models compared to the one-size-fits-all models. The statistical test (one-way ANOVA) also showed differences between the one-size-fits-all model and personalized models (p-value $<$ 0.01). In addition, the comparison between our proposed PA model and the transfer learning method showed that the PA model performed better in terms of the f1-score (paired t-test, p-value $<$ 0.01). Moreover, we found that the PA personalized model size was 68\% smaller than the transfer learning personalized model size.

\begin{table}[]
\caption{Stress detection model performances (f1-score) with different personalized strategies in the modality fusion network (MFN): one-size-fits-all, transfer learning (TL), and personalized attention (PA). PA shows higher f1-scores than TL (paired t-test, p-value $<$ 0.01).
}
\centering
\label{res:pa}
\vspace{-0.2cm}
\begin{tabular}{c|c|c|}
\cline{2-3}
                                                                                              & Personalized Model Size & F1-Score                   \\ \hline
\multicolumn{1}{|c|}{\begin{tabular}[c]{@{}c@{}}MFN \\ (one-size-fits-all)\end{tabular}}      & -                       & 0.693 $\pm$ 0.006          \\ \hline
\multicolumn{1}{|c|}{\begin{tabular}[c]{@{}c@{}}MFN \\ (transfer learning)\end{tabular}}      & 29.3 MB                 & 0.751 $\pm$ 0.010          \\ \hline
\multicolumn{1}{|c|}{\begin{tabular}[c]{@{}c@{}}MFN \\ (personalized attention)\end{tabular}} & \textbf{9.1 MB}         & \textbf{0.774 $\pm$ 0.007} \\ \hline
\end{tabular}
\vspace{-0.5cm}
\end{table}

\vspace{-0.1cm}
\section{Discussion}
\vspace{-0.1cm}
We tackled two common problems in multimodal in-the-wild emotion detection: (i) missing modalities and (ii) model personalization. In previous sections, we showed that MFN can be adaptive to infer stress levels even in the presence of missing modality; and MFN provided better f1-score performance than multiple SAN models. The PA method also showed promising results with higher f1-scores using a smaller number of model parameters (lighter model size) than previous state-of-the-art methods. In this section, we discuss some of the posterior analysis from the results, as well as the practical implications and limitations of this work.

\vspace{-0.1cm}
\subsection{Error Analysis}
\begin{figure}
	\centering
	\includegraphics[scale=0.44]{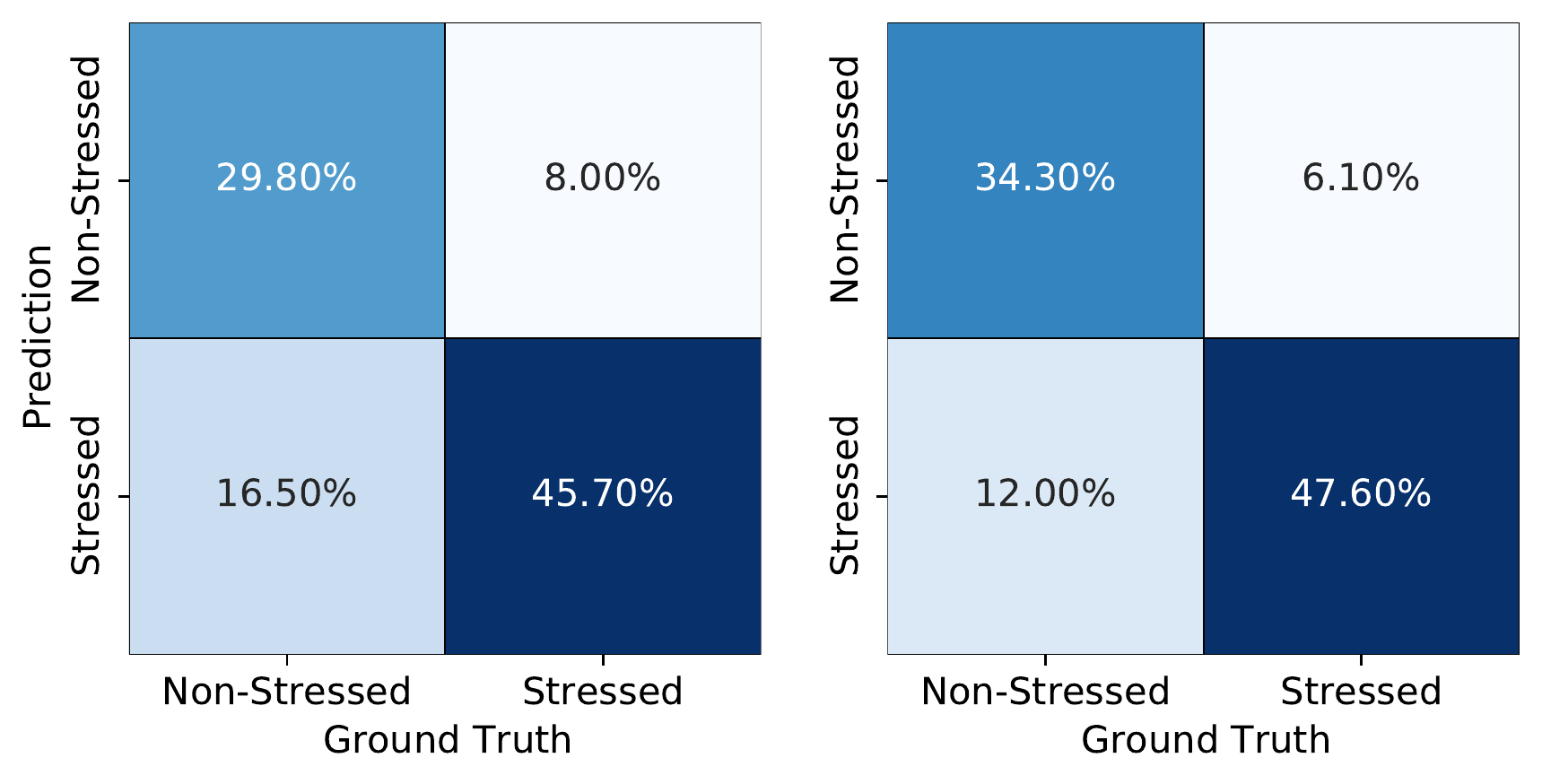}
	\vspace{-0.4cm}
	\caption{Confusion matrix tables of binary stress detection in test set using one-size-fits-all modality fusion network (left) and personalized attention framework (right)}
	\label{fig:cm}
	\vspace{-0.5cm}
\end{figure}

By analyzing the model outputs, we found that detecting stress labels showed higher accuracy than detecting non-stress labels. Such a situation was observed in both the generalized model and the individualized model. One benefit of such a phenomenon is that we can get more accurate estimations when using the model to help people under stress. Figure \ref{fig:cm} shows the confusion matrices for the estimation values of one-size-fits-all MFN model and the PA framework, respectively.  The two confusion matrices showed that the models achieved higher recall scores for stress classification than non-stressed labels, with their recall being 60.8\% and 83.1\%, respectively. In the PA model results, the recall score for non-stressed predictions was improved compared to the generalized model, which increased to 73.5\%. 

Although personalized models improved model performance, there was a difference in the model performance for detecting stress and non-stress labels. Such differences might be caused by label distribution imbalance in the dataset. Although we applied the focal loss method, the label imbalance problem have not been fully solved.

\subsection{Personalized Stress Detection Insights}
We found individual differences in participants' reported momentary stress levels. For example, the maximum, minimum, median percentages of stressed labels across each participant were 100\%, 4.6\%, and 45.7\%, respectively with a standard deviation (STD) value of 31.0\%. To investigate our model performances on individuals, we analyzed the stress detection results from MFN and PA on each subject. In the test results using the generalized model, 8 out of the 41 participants had f1-scores of 85\% or higher, while 6 participants had f1-scores below 50\%. According to our PA framework results, the number of participants with f1-scores above 85\% increased to 15, while only 2 participants had f1-scores below 50\%. 
Among the 15 participants with high f1-scores in stress detection, we found all labels of 4 participants were reported as stressed. This analysis revealed that the model might estimate more accurate stress labels for these participants with stable subject stress patterns. To explore the relationship between the variability in stress labels and our model performance, we calculated the STD of the binary labels for all participants and applied K-Means algorithms to cluster them into two different groups (k-means centroids label STD: 0.125 (N=11) and 0.438 (N=30)) with the highest Silhouette score of 0.698 compared to other numbers (from 2 to 10) of clusters. With the PA framework, we observed that the f1-scores (average: 86.4\%) among participants with lower STD were significantly higher than the f1-scores (average: 73.4\%) among the higher STD group (t-test, p-value $<$ 0.01).
On the other hand, the shifts of label distributions from training to test set might lead to difficulties in estimating stress labels accurately. For example, for the 2 participants who showed low f1-scores below 50\% with the PA framework, we found that the differences in percentages of stressed labels between the training and test sets were both over 40\%. For example, one participant reported 38\% of labels as stressed in the training set; whereas 89\% of labels as stressed in the test set.

\subsection{Interpretability on the Data Time Steps}
\begin{figure}
	\centering
	\includegraphics[scale=0.40]{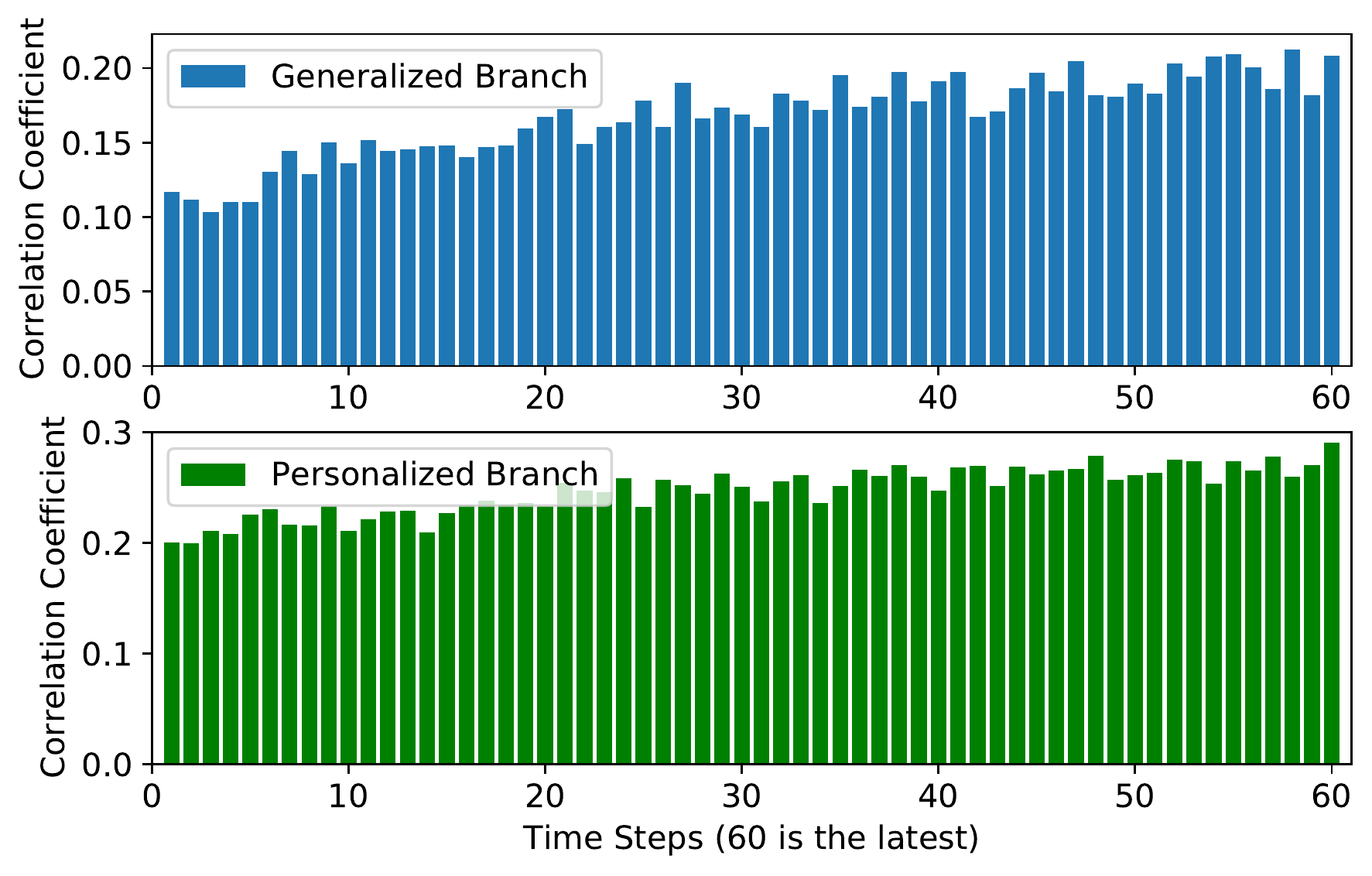}
	\vspace{-0.3cm}
	\caption{The bar plots of average vector correlation coefficients between the input temporal vectors the output of first self-attention block of generalized branch and personalized branch of PA.}
	\label{attention_corr}
	\vspace{-0.5cm}
\end{figure}

To explore the contribution of each time step in the input sequence (60 minutes) to the model, we conducted a correlation analysis for generalized MFN and personalized attention branches in the PA framework, respectively. As the first self-attention blocks in both the generalized branch and personalized branch outputted high-level sequential representations learned from the input, we computed the vector correlations between each time step in the input sequence and the dimensions of intermediate outputs of both branches. Figure \ref{attention_corr} shows the correlations between the input data and model intermediate outputs on each input time step. In both MFN and PA model, we found that as the time step got closer to the time point where the stress was reported, the input data were more correlated with the model output. In addition, we found that the correlations between the input and the output in the personalized branch were higher than those between the input and the generalized branch, which illustrated that the personalized attention for each participant learned more correlated representations than the generalized model.

\subsection{Implications}
\subsubsection{Modality Fusion}
This work provides an insight into using deep learning to solve the incomplete modalities issues in multimodal timeseries sensor data without discarding incomplete data or reconstructing the missing modalities. 
We have proved the effectiveness of MFN with two modalities in this paper. Such logic can be applied to other datasets with more modalities. For example, we can design more input branches to exploit modalities, and we can also use different output branches to estimate  labels under various modalities missing scenarios.

\subsubsection{Personalization} 
Our proposed PA framework can significantly improve stress detection performance compared to generalized one-size-fits-all learning. The design of personalized branches makes it possible to save personalized parameters of the models on user-end edge devices. In real-world applications, we may use the proposed PA framework and distributed computing methods to combine the end-user small models with the server-side one-size-fits-all model to obtain accurate stress estimation for individuals.

 \subsection{Limitations \& Future Work}
 Although this work has made some progress in multimodal wearable sensor-based in-the-wild momentary stress detection, it is undeniable that there are still some limitations. 
First, as mentioned above, there exists the problem of label imbalance. As future work, in addition to trying weight-sensitive loss functions, we will also make new attempts in terms of data sampling, such as oversampling or undersampling to make the training set balanced.
 Second, the diversity of the participants' population could be improved. In the dataset, the age structure of our participants is relatively concentrated on younger age groups, and the gender of the participants is primarily female. In the future, we will collect data from more diverse populations and evaluate the fairness of the model and its adaptability to other users.
 Moreover, generally the deep learning systems perform well but are “black boxes,” and lack insight into the underlying mechanisms. The nonlinear design of neural networks makes it hard to interpret the weights and parameters of networks. At the same time, users may have difficulties trusting the feedback provided by "black boxes". Although understanding deep learning models is still an unsolved problem, in the future, we will extend our current model and bring some interpretability to our stress detection using partially interpretable deep learning design \cite{guo2019exploring, lim2019temporal}.

\section{Conclusion}
We investigated personalized momentary stress estimation using a multimodal wearable dataset where missing data modality was observed. We proposed an MFN structure to adaptively fit data samples and infer stress levels with modality missing conditions. The MFN network shows a 1.6\%  higher f1-score in experiments compared to the performance of the baseline SAN model. Moreover, we designed a PA framework, which individually applied the MFN model to each subject's data. The PA method shows 2.3\% higher f1-score performance with 68.9\% smaller personalized model size than the previous state-of-the-art transfer learning method. Although we achieved some promising results, there are still challenges to overcome. In the future, we will propose new methods in solving our existing issues such as imbalanced data and increasing model interpretability and also test our methods in different datasets.

\newpage
\bibliographystyle{unsrt}
\bibliography{./references}

\begin{thebibliography}{10}

\bibitem{dhabhar2014effects}
Firdaus~S Dhabhar.
\newblock Effects of stress on immune function: the good, the bad, and the
  beautiful.
\newblock {\em Immunologic research}, 58(2-3):193--210, 2014.

\bibitem{aschbacher2013good}
Kirstin Aschbacher, Aoife O’Donovan, Owen~M Wolkowitz, Firdaus~S Dhabhar,
  Yali Su, and Elissa Epel.
\newblock Good stress, bad stress and oxidative stress: insights from
  anticipatory cortisol reactivity.
\newblock {\em Psychoneuroendocrinology}, 38(9):1698--1708, 2013.

\bibitem{kario2003disasters}
Kazuomi Kario, S~McEWEN Bruce, and G~PICKERING Thomas.
\newblock Disasters and the heart: a review of the effects of
  earthquake-induced stress on cardiovascular disease.
\newblock {\em Hypertension Research}, 26(5):355--367, 2003.

\bibitem{khansari1990effects}
David~N Khansari, Anthony~J Murgo, and Robert~E Faith.
\newblock Effects of stress on the immune system.
\newblock {\em Immunology today}, 11:170--175, 1990.

\bibitem{sanchez2017towards}
Wendy Sanchez, Alicia Martinez, and Miguel Gonzalez.
\newblock Towards job stress recognition based on behavior and physiological
  features.
\newblock In {\em International conference on ubiquitous computing and ambient
  intelligence}, pages 311--322. Springer, 2017.

\bibitem{gjoreski2017monitoring}
Martin Gjoreski, Mitja Lu{\v{s}}trek, Matja{\v{z}} Gams, and Hristijan
  Gjoreski.
\newblock Monitoring stress with a wrist device using context.
\newblock {\em Journal of biomedical informatics}, 73:159--170, 2017.

\bibitem{shi-stress}
Yuan Shi, Minh~Hoai Nguyen, Patrick Blitz, Brian French, Scott~P. Fisk,
  Fernando~De la~Torre, Asim Smailagic, Daniel~P. Siewiorek, Mustafa al’Absi,
  Emre Ertin, Thomas Kamarck, and Santosh Kumar.
\newblock Personalized stress detection from physiological measurements.
\newblock 2010.

\bibitem{ma2021smil}
Mengmeng Ma, Jian Ren, Long Zhao, Sergey Tulyakov, Cathy Wu, and Xi~Peng.
\newblock Smil: Multimodal learning with severely missing modality.
\newblock {\em arXiv preprint arXiv:2103.05677}, 2021.

\bibitem{pandey2017variational}
Gaurav Pandey and Ambedkar Dukkipati.
\newblock Variational methods for conditional multimodal deep learning.
\newblock In {\em 2017 International Joint Conference on Neural Networks
  (IJCNN)}, pages 308--315. IEEE, 2017.

\bibitem{wang2018lrmm}
Cheng Wang, Mathias Niepert, and Hui Li.
\newblock Lrmm: learning to recommend with missing modalities.
\newblock {\em arXiv preprint arXiv:1808.06791}, 2018.

\bibitem{jaques2017multimodal}
Natasha Jaques, Sara Taylor, Akane Sano, and Rosalind Picard.
\newblock Multimodal autoencoder: A deep learning approach to filling in
  missing sensor data and enabling better mood prediction.
\newblock In {\em 2017 Seventh International Conference on Affective Computing
  and Intelligent Interaction (ACII)}, pages 202--208. IEEE, 2017.

\bibitem{zhang2019ssim}
Yi-Fan Zhang, Peter~J Thorburn, Wei Xiang, and Peter Fitch.
\newblock Ssim—a deep learning approach for recovering missing time series
  sensor data.
\newblock {\em IEEE Internet of Things Journal}, 6(4):6618--6628, 2019.

\bibitem{peralta2021data}
Maxime Peralta, Pierre Jannin, Claire Haegelen, and John~SH Baxter.
\newblock Data imputation and compression for parkinson's disease clinical
  questionnaires.
\newblock {\em Artificial Intelligence in Medicine}, 114:102051, 2021.

\bibitem{bucur2018early}
Beniamin Bucur, Iulia {\c{S}}omfelean, Alexandru Ghiuru{\c{t}}an, Camelia
  Lemnaru, and Mihaela D{\^\i}n{\c{s}}oreanu.
\newblock An early fusion approach for multimodal emotion recognition using
  deep recurrent networks.
\newblock In {\em 2018 IEEE 14th International Conference on Intelligent
  Computer Communication and Processing (ICCP)}, pages 71--78. IEEE, 2018.

\bibitem{mittal2020m3er}
Trisha Mittal, Uttaran Bhattacharya, Rohan Chandra, Aniket Bera, and Dinesh
  Manocha.
\newblock M3er: Multiplicative multimodal emotion recognition using facial,
  textual, and speech cues.
\newblock In {\em Proceedings of the AAAI Conference on Artificial
  Intelligence}, volume~34, pages 1359--1367, 2020.

\bibitem{can2019continuous}
Yekta~Said Can, Niaz Chalabianloo, Deniz Ekiz, and Cem Ersoy.
\newblock Continuous stress detection using wearable sensors in real life:
  Algorithmic programming contest case study.
\newblock {\em Sensors}, 19(8):1849, 2019.

\bibitem{Previous-cla}
Sara~Ann Taylor, Natasha Jaques, Ehimwenma Nosakhare, Akane Sano, and Rosalind
  Picard.
\newblock Personalized multitask learning for predicting tomorrow's mood,
  stress, and health.
\newblock {\em IEEE Transactions on Affective Computing}, 2017.

\bibitem{saeed2018model}
Aaqib Saeed, Tanir Ozcelebi, Johan Lukkien, Jan~BF van Erp, and Stojan
  Trajanovski.
\newblock Model adaptation and personalization for physiological stress
  detection.
\newblock In {\em 2018 IEEE 5th International Conference on Data Science and
  Advanced Analytics (DSAA)}, pages 209--216. IEEE, 2018.

\bibitem{yu2020passive}
Han Yu and Akane Sano.
\newblock Passive sensor data based future mood, health, and stress prediction:
  User adaptation using deep learning.
\newblock In {\em 2020 42nd Annual International Conference of the IEEE
  Engineering in Medicine \& Biology Society (EMBC)}, pages 5884--5887. IEEE,
  2020.

\bibitem{umematsu2019improving}
Terumi Umematsu, Akane Sano, Sara Taylor, and Rosalind~W Picard.
\newblock Improving students' daily life stress forecasting using lstm neural
  networks.
\newblock In {\em 2019 IEEE EMBS International Conference on Biomedical \&
  Health Informatics (BHI)}, pages 1--4. IEEE, 2019.

\bibitem{Previous-reg2}
Han Yu, Elizabeth~B Klerman, Rosalind~W Picard, and Akane Sano.
\newblock Personalized wellbeing prediction using behavioral, physiological and
  weather data.
\newblock In {\em 2019 IEEE EMBS International Conference on Biomedical \&
  Health Informatics (BHI)}, pages 1--4. IEEE, 2019.

\bibitem{cai2018deep}
Lei Cai, Zhengyang Wang, Hongyang Gao, Dinggang Shen, and Shuiwang Ji.
\newblock Deep adversarial learning for multi-modality missing data completion.
\newblock In {\em Proceedings of the 24th ACM SIGKDD International Conference
  on Knowledge Discovery \& Data Mining}, pages 1158--1166, 2018.

\bibitem{smets2018towards}
Elena Smets.
\newblock Towards large-scale physiological stress detection in an ambulant
  environment.
\newblock 2018.

\bibitem{selvin2017stock}
Sreelekshmy Selvin, R~Vinayakumar, EA~Gopalakrishnan, Vijay~Krishna Menon, and
  KP~Soman.
\newblock Stock price prediction using lstm, rnn and cnn-sliding window model.
\newblock In {\em 2017 international conference on advances in computing,
  communications and informatics (icacci)}, pages 1643--1647. IEEE, 2017.

\bibitem{cho2014learning}
Kyunghyun Cho, Bart Van~Merri{\"e}nboer, Caglar Gulcehre, Dzmitry Bahdanau,
  Fethi Bougares, Holger Schwenk, and Yoshua Bengio.
\newblock Learning phrase representations using rnn encoder-decoder for
  statistical machine translation.
\newblock {\em arXiv preprint arXiv:1406.1078}, 2014.

\bibitem{goel2014polyphonic}
Kratarth Goel, Raunaq Vohra, and Jajati~Keshari Sahoo.
\newblock Polyphonic music generation by modeling temporal dependencies using a
  rnn-dbn.
\newblock In {\em International Conference on Artificial Neural Networks},
  pages 217--224. Springer, 2014.

\bibitem{wang2016survey}
Feng Wang and David~MJ Tax.
\newblock Survey on the attention based rnn model and its applications in
  computer vision.
\newblock {\em arXiv preprint arXiv:1601.06823}, 2016.

\bibitem{hori2017advances}
Takaaki Hori, Shinji Watanabe, Yu~Zhang, and William Chan.
\newblock Advances in joint ctc-attention based end-to-end speech recognition
  with a deep cnn encoder and rnn-lm.
\newblock {\em arXiv preprint arXiv:1706.02737}, 2017.

\bibitem{vaswani2017attention}
Ashish Vaswani, Noam Shazeer, Niki Parmar, Jakob Uszkoreit, Llion Jones,
  Aidan~N Gomez, Lukasz Kaiser, and Illia Polosukhin.
\newblock Attention is all you need.
\newblock {\em arXiv preprint arXiv:1706.03762}, 2017.

\bibitem{so2019evolved}
David So, Quoc Le, and Chen Liang.
\newblock The evolved transformer.
\newblock In {\em International Conference on Machine Learning}, pages
  5877--5886. PMLR, 2019.

\bibitem{he2016deep}
Kaiming He, Xiangyu Zhang, Shaoqing Ren, and Jian Sun.
\newblock Deep residual learning for image recognition.
\newblock In {\em Proceedings of the IEEE conference on computer vision and
  pattern recognition}, pages 770--778, 2016.

\bibitem{liu2019nrpa}
Hongtao Liu, Fangzhao Wu, Wenjun Wang, Xianchen Wang, Pengfei Jiao, Chuhan Wu,
  and Xing Xie.
\newblock Nrpa: neural recommendation with personalized attention.
\newblock In {\em Proceedings of the 42nd International ACM SIGIR Conference on
  Research and Development in Information Retrieval}, pages 1233--1236, 2019.

\bibitem{lin2017focal}
Tsung-Yi Lin, Priya Goyal, Ross Girshick, Kaiming He, and Piotr Doll{\'a}r.
\newblock Focal loss for dense object detection.
\newblock In {\em Proceedings of the IEEE international conference on computer
  vision}, pages 2980--2988, 2017.

\bibitem{guo2019exploring}
Tian Guo, Tao Lin, and Nino Antulov-Fantulin.
\newblock Exploring interpretable lstm neural networks over multi-variable
  data.
\newblock {\em arXiv preprint arXiv:1905.12034}, 2019.

\bibitem{lim2019temporal}
Bryan Lim, Sercan~O Arik, Nicolas Loeff, and Tomas Pfister.
\newblock Temporal fusion transformers for interpretable multi-horizon time
  series forecasting.
\newblock {\em arXiv preprint arXiv:1912.09363}, 2019.

\end{thebibliography}

\vspace{12pt}

\end{document}